\documentclass[acmlarge]{acmart}

\AtBeginDocument{%
	\providecommand\BibTeX{{%
			\normalfont B\kern-0.5em{\scshape i\kern-0.25em b}\kern-0.8em\TeX}}}

\acmJournal{TOMM}
\acmVolume{01}
\acmNumber{01}
\acmArticle{01}
\acmMonth{06}

\input{preamble}

\usepackage{cleveref}

\begin{document}

\title{SETTI: A Self-Supervised Adversarial Malware Detection Architecture in an IoT Environment}
\author{Marjan Golmaryami}
\affiliation{%
	\institution{Computer Engineering and Information Technology Department, Shiraz University of Technology}
	\city{Shiraz}
	\country{Iran}}
\email{m.golmaryami@sutech.ac.ir}

\author{Rahim Taheri}
\affiliation{%
	\institution{King’s Communications, Learning and Information Processing (kclip) lab, King's College London}
	\city{London}
	\country{United Kingdom}
}
\email{rahim.taheri@kcl.ac.uk}

\author{Zahra Pooranian}
\affiliation{%
	\institution{5GIC \& 6GIC, Institute for Communication Systems (ICS), University of Surrey}
	\city{Guildford}
	\country{United Kingdom}}
\email{z.pooranian@surrey.ac.uk}

\author{Mohammad Shojafar}
\authornote{This is the corresponding author}
\authornote{All authors contributed equally to this research.}

\affiliation{%
	\institution{5GIC \& 6GIC, Institute for Communication Systems, University of Surrey}
	\city{Guildford}
	\country{United Kingdom}}
\email{m.shojafar@surrey.ac.uk}

\author{Pei Xiao}
\affiliation{%
	\institution{5GIC \& 6GIC, Institute for Communication Systems, University of Surrey}
	\city{Guildford}
	\country{United Kingdom}}
\email{p.xiao@surrey.ac.uk}

\renewcommand{\shortauthors}{Golmaryami, Taheri and Pooranian, et al.}

\begin{abstract}
In recent years, malware detection has become an active research topic in the area of Internet of Things (IoT) security. The principle is to exploit knowledge from large quantities of continuously generated malware. Existing algorithms practise available malware features for IoT devices and lack real-time prediction behaviours. More research is thus required on malware detection to cope with real-time misclassification of the input IoT data. Motivated by this, in this paper we propose an adversarial self-supervised architecture for detecting malware in IoT networks, SETTI, considering samples of IoT network traffic that may not be labeled. In the SETTI architecture, we design \emph{three} self-supervised attack techniques, namely \emph{Self-MDS}, \emph{GSelf-MDS} and \emph{ASelf-MDS}. The Self-MDS method considers the IoT input data and the adversarial sample generation in real-time. The GSelf-MDS builds a generative adversarial network model to generate adversarial samples in the self-supervised structure. Finally, ASelf-MDS utilizes \emph{three} well-known perturbation sample techniques to develop adversarial malware and inject it over the self-supervised architecture. Also, we apply a defence method to mitigate these attacks, namely \emph{adversarial self-supervised training} to protect the malware detection architecture against injecting the malicious samples. To validate the attack and defence algorithms, we conduct experiments on two recent IoT datasets: IoT23 and NBIoT. Comparison of the results shows that in the IoT23 dataset, the Self-MDS method has the most damaging consequences from the attacker's point of view by reducing the accuracy rate from 98\% to 74\%. In the NBIoT dataset, the ASelf-MDS method is the most devastating algorithm that can plunge the accuracy rate from 98\% to 77\%.
\end{abstract}


\maketitle

\section{Introduction}\label{sez:1}
Machine learning (ML) has become a powerful tool for identifying and predicting system behaviour in recent years. ML has been widely used in security applications such as identity recognition/verification and fraud detection programs~\cite{li2020deep}. However, these programs are vulnerable to adversarial attacks and malicious behaviour. Several lines of research indicate that ML models can be easily exposed to deceptions, especially in the case of precisely crafted perturbations to the input training sample, mainly targeting at
the features. For example, the entire original sample can be viewed by an adversary who then adds a perturbation to some feature of the sample as well as to the remaining unobserved samples. Adding the perturbation is particularly critical in dynamic environments such as computer networks.

The majority of recent adversarial poisoning ML algorithms need to observe the entire original data sample, which is inserted into the target model to modify any part of the sample, such as speech adversarial poisoning ML algorithms. In these adversarial poisoning ML algorithms, they first develop the perturbation model, add the new instances to the original dataset, and finally input the new dataset (original sample and perturbation samples) into the speech recognition system.

\textcolor{black}{To cope with these issues, several solutions have emerged in literature~\cite{sadeghzadeh2021adversarial,saeed2020federated,yu2020you}. Nevertheless, those solutions have certain limitations. For example, the model is unfeasible when the target system mainly relies on continuously processed streaming input in network traffic~\cite{cruz2016cybersecurity}. To prevent poisoning attacks, we need to have new solutions that can simultaneously process the adversary behaviours in real-time, regulate the additional perturbations in the dataset, and rely on the final conclusion based on the whole dataset samples. In the following, we showcase some of the promising solutions that works in this way.}

\begin{itemize}[leftmargin=*]
\item \emph{Network Intrusion Detection Systems.} Through continuous monitoring and observation of network behaviors, computer networks employ automatic ML algorithms. An adversary can intervene in the final automatic ML model in an IDS by accurately perturbing the relevant traffic templates. Thanks to the IDS that decides based on the long sequence of observations in the network's specific period, it is difficult for an adversary to subvert the network's historical samples unless the adversary can influence future traffic behavior with some strategies such as traffic manipulations.

\item \emph{Real-time Speech Processing Systems.} The speech processing system also uses ML for real-time applications like security-sensitive services. For example, an adversary could manipulate the system output by modifying the designed noise model, mostly outside of the human ear's bounds, which is easy to discover when the human voice spreads through the air. Therefore, the adversary can only generate noise signals based on past speech signals and superimpose the noise only on future speech signals. In contrast, the speech processing system will complete its task using the entire speech segment. In this case, the adversary has a trade-off between observation and action space. To elaborate on this, let us denote $S$ as the input sequence of a target system, and the adversary could initially create the related adversarial perturbation samples. Nevertheless, suppose the adversary does not see the sequence $S$. In this case, it can add malicious samples to any part of the input sequence $S$, we say \emph{the adversary has maximum action space with minimum observation}. The adversary, on the other hand, has access to the entire sequence $S$ if it adds the adversarial perturbation samples at the end. However, there is no possibility to add perturbations to the data, we say \emph{the adversary to have maximum observation with minimum action space}. In comparison, the former has difficulty obtaining an optimal perturbation for $S$ without having any access to the sequence, while the latter cannot implement the attack at all.
\end{itemize}

\subsection{Motivation of the Paper}
Self-supervised learning (SSL) methods are cutting-edge mechanisms~\cite{gidaris2018unsupervised,zou2018unsupervised,dosovitskiy2015discriminative} that are widely used in real-time applications like video applications. When the labelled data are few, using SSL increases the representation rounds~\cite{qin2020sp}. However, SSL solutions lack an accurate accuracy ratio, especially for fully-supervised learning (FSL) algorithms. Therefore, most recent efforts are mainly devoted to targeting performance enhancement. Besides, the already designed models are effective for the SSL and fully labelled datasets. To cope with this limitation, in this paper, we combine the SSL with the fully labelled Internet of Things (IoT) dataset to enhance the model's robustness against the adversarial examples, especially while concurrently monitoring the accuracy and performance of the ML model.

\textcolor{black}{We have also discovered that SSL solutions can improve a long-standing and underexplored problem that are mainly targeted at different distributions. Also, SSL methods can significantly enhance robustness ~\cite{schmidt2018adversarially,kurakin2016adversarial}. In a nutshell, we can achieve sensible improvements in adversarial robustness, label corruption, and common input corruption with SSL. Hence, some questions arise that we aim to address in this study:
(i) Can we improve the design of an architecture addressing SSL on the malware dataset? (\Cref{architecture})
(ii) How can we design an attack mechanism to craft the ML model using SSL deliberately? (\Cref{Self-MDS} to \Cref{AdversarialSelf-Supervised})
(iii) Can we introduce a robust defence solution tackling the real-time SSL attack on the crafted model in an adversarial malware detection system? (\Cref{AdversarialSelf-Defence}) 
In the rest of the paper, we delineate the response to these queries.}

\subsection{The Main Goal and Contributions}  
This paper aims to design new self-supervised adversarial attacks by overcoming issues in streaming inputs to achieve robustness in the malware detection system. We develop several attack techniques to use continuously observed data to approximate an optimal adversarial perturbation for future time points using a self-supervised learning architecture to address this issue. This paper also proposes a real-time adversarial attack method for ML-based malware detection with streaming inputs. In brief, the essential contributions of the paper are as follows:
    \begin{itemize}[leftmargin=*]
    \item First, we present a robust self-supervised based architecture for network malware detection systems, named \emph{SETTI}, including real-time streaming inputs.
    \item Second, We propose \emph{three} adversarial attacks based on self-supervised learning to conduct network malware floating on benign data samples.
    \item Next, We present a self-supervised defence mechanism against the proposed attack to a robust network malware detection system.
    \item Finally, we perform experiments and validate the attacks and defence mechanisms on \emph{two} IoT datasets using different features.
     \end{itemize}

  \noindent\textbf{Roadmap.} The remainder of the paper is organised as follows. We give an exhaustive study of the state-of-the-art in Section~\ref{relatedWork}. Section~\ref{Preliminaries} explains the data stream network malware representation. Section~\ref{problemDefinition} describes our proposed architecture, attacks and defences. Section~\ref{resultAnalysis} presents the performance evaluation. Finally, Section~\ref{conclusion} summarises the paper and gives outlooks.

\section{Related Work}\label{relatedWork} 
We begin this section by discussing the related robustness of self-supervised approaches in Section~\ref{sec:SS-related} and the robustness of real-time malware detection methods in Section~\ref{sec:RL-related}. To aid in comprehension, the comparison results of techniques are presented in Table~\ref{table: comparison_table}, which indicates that there do not exist any self-supervised approaches designed for the adversarial malware detection system.

\begin{table}
\caption{\small Comparison between different self-supervised (SS) representation learning algorithms. RT:= Real-Time; GN:= Generator. Advr:= Adversarial. IDS:= Intrusion Detection System; NLP:= Natural Language Processing;  }
\label{table: comparison_table}
\begin{center}
\begin{adjustbox}{max width=\textwidth}
\begin{tabular}{lcccccc}
\hline
 \rowcolor[HTML]{EFEFEF}  \multirow{1}{*}{\textbf{Ref.}}&
  \textbf{Field}& \textbf{SS} & \textbf{GN}&\textbf{Pretask}&\textbf{Advr}&\textbf{RT}\\ \hline
  \textbf{\cite{joshi2020spanbert}}&NLP&$\checkmark$&$\checkmark$&Masked language&$-$&$-$\\
  
  \textbf{\cite{kim2018learning}}&CV&$\checkmark$&$\checkmark$&Inpainting&$-$&$-$\\
  \textbf{\cite{kingma2018glow}}&CV&$\checkmark$&$\checkmark$&Image reconstruction&$-$&$\checkmark$\\
  
  \textbf{\cite{clark2020electra}}&NLP&$\checkmark$&$\checkmark$&Masked words&$\checkmark$&$-$\\
  
  \textbf{\cite{ding2018semi}}&Graph&$\checkmark$&$\checkmark$&Node classification&$\checkmark$&$-$\\

  \textbf{\cite{dai2018adversarial}}&Graph&$\checkmark$&$\checkmark$&Link prediction&$\checkmark$&$-$\\

  \textbf{\cite{donahue2019large}}&CV&$\checkmark$&$\checkmark$&Image reconstruction&$\checkmark$&$-$\\

   \textbf{SETTI}&\textbf{IDS}&$\checkmark$&$\checkmark$&\textbf{Malware detection}&$\checkmark$&$\checkmark$\\
  \hline
\end{tabular}
\end{adjustbox}
\end{center}
\end{table}

\subsection{Self-supervised Learning (SSL) Approaches} \label{sec:SS-related}
\textcolor{black}{Several SSL techniques have been proposed in the literature, each exploring different pretext activities for image and healthcare applications~\cite{alizadehsani2021uncertainty}. The study in \cite{doersch2015unsupervised} estimates the relative position of image patches and applies the resulting representation to enhance object detection. The method in \cite{dosovitskiy2015discriminative} creates classification algorithm training by transforming seed image patches. Some other technologies utilise colorization as a proxy task \cite{larsson2017colorization}, deep clustering methods~\cite{alwassel2020self}, and methods that maximise mutual information \cite{hjelm2018learning} with high-level representations~\cite{oord2018representation,henaff2020data,ghasedi2018semi}. These studies address the utility of SSL. However, they do not consider labelling data and its effect on robustness, which is addressed in this paper.}

Improving model robustness helps ensure the resilience of ML models across various inadequate training and testing conditions. In~\cite{hendrycks2019benchmarking}, the authors evaluate the designed ML models with real-world image corruptions and a large set of distortions. Other robust ML algorithms may also be able to run over corrupted labels~\cite{nettleton2010study,patrini2017making}.The Gold Loss Correction (GLC) method in~\cite{hendrycks2018using} utilizes a small set of trusted labels to improve the setting's accuracy. In the presence of high degrees of label corruption, the ML models overfit false samples into the corrupted labels, such as~\cite{zhang2018generalized,hendrycks2019using,arshad2018samadroid,zhang2021self}. These algorithms could suggest extra unsupervised objective training with reliable signals. A projected gradient descent (PGD) adversarial training algorithm was developed in~\cite{madry2017towards} to confirm adversarial robustness. Their proposed algorithm trains to maximise the robustness of the ML model. In~\cite{zhang2019theoretically}, the authors design an enhanced method using a modified loss function and disclose the trade-off between accuracy before and after an adversarial attack.

\subsection{Real-time Approaches}\label{sec:RL-related}
In recent decades, real-time detection systems have been developed to monitor sensitive activities such as file locations and their associated addresses. Nevertheless, most methods only focus on particular features. In contrast, malware execution involves a set of behaviour chains such that attackers could alter features or sensitive behaviours with the traditional ones. By doing so, the malware can pass the detection system. Many anti-detection methods use the above schemes, and in contrast, detectors could only increase the number of features to temporarily address the problem. Further, some malware, including encryption ransomware, would conceive their behaviours with dynamic encryption/decryption or hook/injection methods and would perform fundamental behaviour at the last moment~\cite{bazrafshan2013survey}. When detection systems discover anomalies, it is too late to prevent malicious behaviour effects, and attackers have already fulfilled their goals~\cite{xia2021sedmdroid}.

If we want to consider other areas besides malware analysis, in the image processing area, the well-known works~\cite{bouguet2001pyramidal,menze2015object} introduce a tool based on optical flow to analyse moving objects in video files. It has been utilised to discover the matching between incoming image frames and self-supervised learning (SSL)~\cite{zhang2020empowering}. An SSL solution is proposed in~\cite{lieb2005adaptive} for road tracking. They adopt an optical flow mechanism to detect road templates for matching frames and control road segmentation. This approach does not require any method to use the information on appearance or the pavement structure. The authors of \cite{yang2019unsupervised} introduce an approach to detect the moving objects in an image. Their approach includes two components: a generator and an inpainter. The generator component produces a set of objects in an image with the optical flow information. The second component re-estimates the optical flow of that set. \\

In malware, the call sequences of malware have a specific purpose. They can be used to analyse malware behavior, as demonstrated in~\cite{alqurashi2017comparison,dib2022evoliot}. Although some anti-detection techniques inject some obfuscated services to hide malware's purpose, in the execution phase, its critical calls that refer to its real goals must be invoked. In a nutshell, using the call sequence in real-time detection is a useful prototype. Over the last decade, traffic analysis has been the focus of most research in real-time detection. However, a few real-time methods are available to detect malware, and most of the research has been devoted to static detection methods. In the following, we review some related detection methods. In~\cite{chen2016statistical}, the authors developed a real-time detection system for Twitter's spam. They first extract statistical features from spam and normal mail and then adopt a machine learning algorithm consisting of two layers: a learning layer and a training layer. The former is used to discover the alternation for unlabeled spam tweets. While the latter receives the changed features. The training layers can also learn new features in the training step. The results confirm the improvement in the system's accuracy in spam detection rate in real-time. Although the work focused on real-time spam detection, the idea can be applied to real-time malware detection. The study referred to in~\cite{naval2015employing} presented a protection approach for real-time ransomware detection. The method assessed the operation of the system and then adopted access control on essential functions. Besides, the results demonstrate the promising performance of the method for detecting and preventing ransomware. However, the energy consumption of their method for the CPU usage is high (around 70\%), and it consumes a huge amount of memory, which is around 2,5GB. Finally, the method is only applicable to powerful servers, and it is not applicable as a general-purpose service for mobile apps.
DNA-Droid~\cite{gharib2017dna} is a framework to detect ransomware, consisting of static and dynamic. The authors claimed that the proposed method reached an accuracy of 98\% in detecting malware than other existing methods. The point that has always been considered in various studies \cite{taheri2020fed} is that the accuracy of detection methods after using the defense method is still lower than the accuracy of the proposed method in the case that an attack did not occur.

There is also an existing traditional optical flow method \cite{lucas1981iterative}. Some existing solutions predict optical flows using a self-supervised/unsupervised (SSL/SUL) neural network~\cite{meister2018unflow}, \cite{alletto2018self}, \cite{guo2020coverage}. For example, in \cite{mahendran2018cross}, the authors develop the SSL of a deep CNN over self-supervision through the image files. Later on, the authors in \cite{liu2019selflow} design an SSL unblocked optical flow information to study the optical occlusion flow. An SUL is presented in \cite{ren2017unsupervised} to predict the optical flow. The study in \cite{yin2018geonet} designs an SUL GeoNet to jointly estimate the monocular depth map on set of real-time image files. 
The authors of ~\cite{mugnai2022fine} use Semi-Supervised Learning to increase the quantity of training data available and to improve the performance of Fine-Grained Visual Categorization. Their approach employs unlabeled data and an adversarial optimization strategy in which a second-order pooling model is used to generate the internal feature representation. This combination enables backpropagation of component information, represented by second-order pooling, onto unlabeled input in an adversarial training setting. In~\cite{song2022self}, the authors suggested a Self-supervised Heterogeneous Graph Network that could model the interactions between elements and the calories in meals at the same time. Users and features are put together in a heterogeneous network, and directed edges show how they interact with each other. Then, they use self-supervised prediction to look at the connection between components. Finally, they emphasize that the user's choice is important.

As proposed by the authors~\cite{duan2022novel}, MsGM is a multi-sample generation model for black-box model attacks that uses a large number of samples. Multi-sample generation, replacement model training, and adversarial sample generation and attack are primarily made up of three parts. To begin with, they develop a multi-task generation model to learn the distribution of the original dataset, which is then used to train the model. A random signal with a given distribution is converted into the common features of an original dataset by deconvolution procedures, and then many identical sub-networks are generated according to varied input circumstances, resulting in the corresponding targeted samples. Then, the generated sample features produce a variety of different outputs after being queried by the black-box model and trained by the substitute model, which is then used to construct a variety of different loss functions in order to optimise and update both the generator and substitute model, respectively. Once this has been accomplished, various commonly used white-box attack methods are employed to attack the replacement model in order to create appropriate adversarial samples, which are then employed to target the black-box model. None of the above work has provided adversarial algorithms for self-supervised methods when few data points are available. They also don't take into account real-time methods for learning on your own, and this paper shows how it works in both cases.

Unlike the solutions mentioned above, in this paper, we apply self-supervised training algorithms to estimate the next representations of the data points and then we build a loss function by these points to refine the training of our convolutional neural network (CNN) classification algorithm. This paper also discusses new methods for adversarial attacks and defenses against such attacks.

\section{Preliminaries}\label{Preliminaries}
In this section, we provide a detailed description of the data stream for the network malware dataset. Denote $x=\{x_1,x_2,x_3,\ldots x_n\} \in R^{m\times n}$ as a time-series network malware dataset that has $n$ points and each point is $x_i\in R^m $. $f: R^{m\times n}\rightarrow \{y_1,y_2,y_3,\ldots y_c\}$ is a classification algorithm that maps each time-series instance, such as $x$, to a discrete label set. The adversary's purpose is to design a function to generate real-time perturbations to continuously use the observed data from $x$ to achieve optimal adversarial perturbation.

Suppose that after observing the $t$ samples, the adversary uses the function $p(.)$ to add perturbation to the dataset. In the process of adding operation, another $k$ samples also imported to the dataset. In this case, the perturbation value denoted by $\lambda$ is calculated as follows:
\begin{equation}
\label{eq:eq1}\small
   \lambda_t=
\begin{cases}
p({\{x_1,x_2,x_3,\ldots x_{k-1}\}}), & x_k<t<=n \\
0 & \text{otherwise}
\end{cases}
\end{equation}
Here, we define a metric $m(.)$ to measure adversarial perturbation's perceptibility. Then, our goal is to solve the following optimization problem for adversarial attacks.

\begin{equation}
\label{eq:eq2}
\begin{cases}
\textbf{Minimize}\quad m(\lambda={\lambda_1,\lambda_2,\ldots,\lambda_n})\\
s.t.\quad\quad\qquad f(x+\lambda)\neq f(x) 
\end{cases}
\end{equation}

Equation~\eqref{eq:eq1} implies that adversarial perturbation is only based on the observed part of the data samples and can only be applied to the samples' unseen part. Even without the limitations of Eq.~\eqref{eq:eq1}, the direct solution of Eq.~\eqref{eq:eq2} is impossible when $f$ is a deep neural network due to its non-convex nature.

\section{Self-Supervised Malware Detection} \label{problemDefinition}

Self-supervised Learning by rote is an unsupervised learning technique that involves creating a supervised learning task from unlabeled input data. Self-supervised is based on this motivation, first learning useful representations of data from the unlabeled data set and then setting up the representations with a few labels for the supervised downstream part. In this section, we describe our self-supervised malware detection model and related explanations (\Cref{architecture}). Then, we explain our proposed algorithms (\Cref{Self-MDS}-\Cref{AdversarialSelf-Supervised}). Finally, we provide the computational complexity of the attack and defence algorithms (\Cref{TimeComplexity}).

\begin{figure}[!ht]
\centering 
\includegraphics[width=0.69\textwidth]{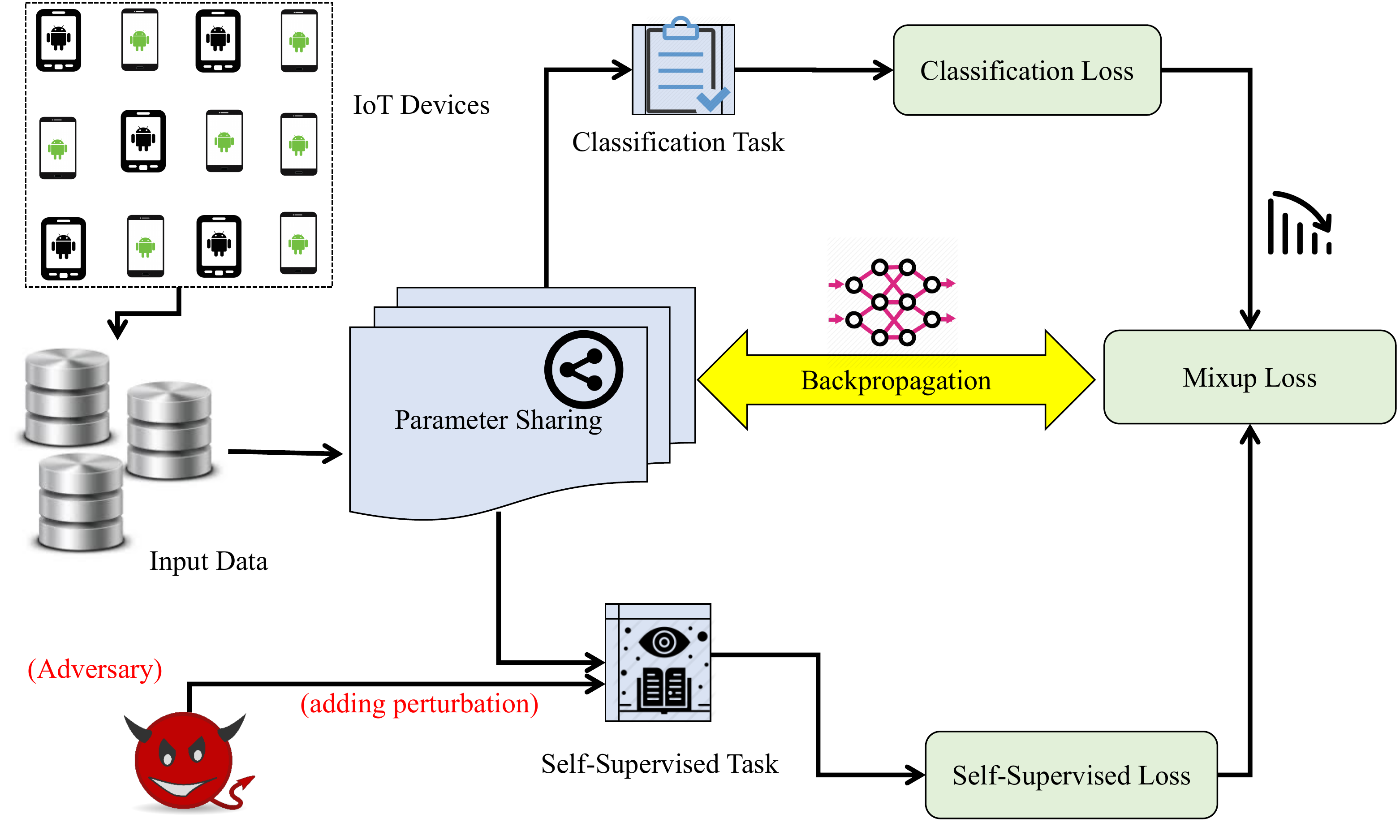}
\caption{\small SETTI: The proposed \underline{S}elf-supervised adv\underline{E}rsarial malware de\underline{T}ection archi\underline{T}ecture for \underline{I}oT malware detection in the presence of an adversary.}
\label{fig:architecture}
\end{figure}

\subsection{SETTI: Proposed Architecture}\label{architecture}
Fig.~\ref{fig:architecture} presents the architecture of the proposed self-supervised adversarial malware detection scheme, SETTI. As can be seen from this figure, the data (including malware and benign) is collected from an environment consisting of IoT devices. In the process of data collection, for various reasons, the data label may not be available or perturbed to the data with an adversary's presence. In Fig.~\ref{fig:architecture}, we present the structure of self-supervised architecture and the way to update the relevant parameters. As can be seen in the self-supervised task section, we use a particular task that is subject to adversarial attacks due to applying this method (see the self-supervised task component in Fig.~\ref{fig:architecture}). This paper also suggests attacks can be of three types that we explain in the following sections.

 \subsection{Self-MDS: Self-Supervised Malware Detection System}\label{Self-MDS}

SSL algorithms form pretext tasks to generate labels with no extra manual annotations, such as  Generative Adversarial Network(GAN)~\cite{goodfellow2014generative}. We apply Self-Supervised Malware detection system based on GAN and then we proposed Self-MDS Attack against it. In the applied Self-Supervised method, our generated labels preserve self-supervision for the network and provide robust representation. In this algorithm, the model is pre-trained by proposed network in~\cite{deng2009imagenet}, like~\cite{doersch2015unsupervised,hung2019scops}. In applied Self-Supervised, we adopt weakly supervised classification algorithm to help weak supervisions and practices weak labels to guide network training, like bounding boxes~\cite{dai2015boxsup,khoreva2017simple}. Since there is a large gap between the fully and weakly supervised malware detection, we create our pretext task based on an ideal detection function equivalence to assist the self-supervision strategy and enhance learning feature-level annotations.

\textcolor{black}{In Self-supervised learning method, first, we train our weakly supervised classification algorithm. Then, we design our self-attention procedure to narrow the supervision gap between fully and weakly supervised learning.This procedure supports network approximation capability.Let $x$ and $y$ be input and output features, with spatial position index $i$ and $j$.}
\textcolor{black}{We define the normalized output signal as $K(x_i)=\sum_{\forall j} w(x_i,x_j)$. In Eq.~\eqref{eq:eq4}, $R(x_j)$ is an aggregation algorithm for input signal $x_j$ and  $W(x_i,x_j)$ is the similarity weights of spatial position index $i$ and $j$ for input $x$. To preserve consistent classification, we generate our self-attention algorithm using equivalent regularization. Then, we define the self-attention approach as~\cite{wang2020self}:}

\begin{equation}
\label{eq:eq4} 
x_{per}=\frac{1}{K(x_i)}\sum w(x_i,x_j)\cdot R(x_j)+ x_i
\end{equation}

Algorithm~\ref{alg:SelfMDS} presents our proposed Self-MDS attack to applied Self-Supervised method. This algorithm consists of a $while$ loop illustrated in lines~\ref{lbl:while-do1} to~\ref{lbl:while-end1}. In this way, we add the entered real-time input samples $x$, and the corresponding label for each sample computed in Eq.~\eqref{eq:eq4}. Then, we use a gradient function and compute an adversarial sample (see lines~\ref{lbl:gradient-start} to~\ref{lbl:gradient-end}). This algorithm designs an approach to receive real-time data and generating adversarial samples.

\begin{algorithm}[t]
	\footnotesize
\caption{\small Self-MDS Attack}
\label{alg:SelfMDS}
\textbf{Input:}$X_{train}$
\textbf{output:} $X_{per}$
\begin{algorithmic}[1]
\color{black}
\footnotesize
\While{true}\label{lbl:while-do1}
\State{$x\leftarrow$  New sample}
\State{$X_{train}=X_{train}+x$}
\If{$x$ is matched with previous samples }
\State{$O\leftarrow x$}
\Else
\State{$S\leftarrow x$}
\EndIf
\If{$0 < length(O)$}
\For{each $i$ in $O$}
\State{Compute gradients $p_t =\nabla_{x} \mathcal{L}(x_{i}, X_{train})$}\label{lbl:gradient-start}
\State{$x_{new} = x_i + \kappa \cdot sign(p_t)$}
\State{$X_{per}=X_{per}+x_{new}$}\label{lbl:gradient-end}
\EndFor
\EndIf
\If{$length(O)==0$}
\State{\textbf{break}}
\EndIf
\EndWhile\label{lbl:while-end1}
\State{\textbf{Return}: $X_{per}$}
\end{algorithmic}
\end{algorithm}
\subsection{GSlef-MDS: GAN-based Self-Supervised Malware Detection System}\label{defence}

In this part, we explain our proposed GAN-based Self-Supervised malware detection system named \emph{GSlef-MDS}. This method adopts GAN, a form of unsupervised learning, and applies labelling over the training dataset. Hence, the GAN method categorizes the sample data into the `real' and `fake' labelled classification. Our approach generates labels from features by focusing on the unlabeled samples and labelling data and supervising them. 

In our proposed method, we design a loss function to control the training phase. It helps to realise the meaningful intermediate representations. Thus, the intermediate terms could store the related structural information for the generated model. GSlef-MDS quickly generates precise labels on the unlabeled data. Since we use GAN in this method, the generated labels achieve high accuracy, as will be evidenced in the result section. During the training phase in GSlef-MDS, the generator and discriminator are in a non-stationary state; hence, the following equation persists:  

\begin{equation}
\label{eq:eqGAN}
\begin{matrix}
V(\mathbb{G},\mathbb{D})=E_{x\sim P_{data(x)}}[\log P_{\mathbb{D}}(S=1|x)] 
+ E_{x\sim P_{\mathbb{G}(x)}}[1-\log P_{\mathbb{D}}(S=0|x)]
\end{matrix}
\end{equation}

In Eq.~\eqref{eq:eqGAN}, $\mathbb{G}$ is generator and $\mathbb{D}$ is discriminator function. The value of function $P_{\mathbb{G}(\cdot)}$ changes during the non-stationary training phase and indicates the distribution of the generated samples in the malware detection system. The formula also shows that training could change the value of $\mathbb{G}$ and $\mathbb{D}$ as gradient descent parameters for the establishment of the learning model.

In this equation, the parameter $\mathbb{D}$ is used to detect the added perturbation to the sample features on the benign sample. In contrast, parameter $\mathbb{G}$ is used to generate a fake sample such that it looks at real/benign sample. In practice, we design $\mathbb{D}$ using two layers in GAN generator function.

In this way, we develop first layer to create change detection, $Q_{\mathbb{D}}$, and the second layer to design the distribution of discrimination samples, $P_{\mathbb{D}}$. Hence, we have:

\begin{equation}
\label{eq:eqGANpro}
\begin{matrix}\small 
L(\mathbb{G})=-V(\mathbb{G},\mathbb{D})-\alpha E_{x\sim P_{\mathbb{G}}}E_{r\sim R}[\log Q_{\mathbb{D}}(R=r|x^r)]\\
L(\mathbb{D})=V(\mathbb{G},\mathbb{D})-\beta E_{x\sim P_{data}}E_{r\sim R}[\log Q_{\mathbb{D}}(R=r|x^r)]
\end{matrix}
\end{equation}
where, $V$ expresses the adversarial criterion, $R$ stands for the set of possible changes, $r$ signifies the chosen change, $x^r$ is the rotated real malware sample, and $\alpha$ and $\beta=1$ are the hyperparameters. Here, parameter $\mathbb{G}$ learns to generate malware sample, whose representations in $\mathbb{D}$'s feature space allow detecting changes. The variable $\alpha$ takes its value within the following range: $0\leq\alpha\leq 1$. This value helps the generated distribution in Eq.~\eqref{eq:eqGANpro} converge to a positive value.

\begin{figure}[!ht]
\centering 
\includegraphics[width=0.69\textwidth]{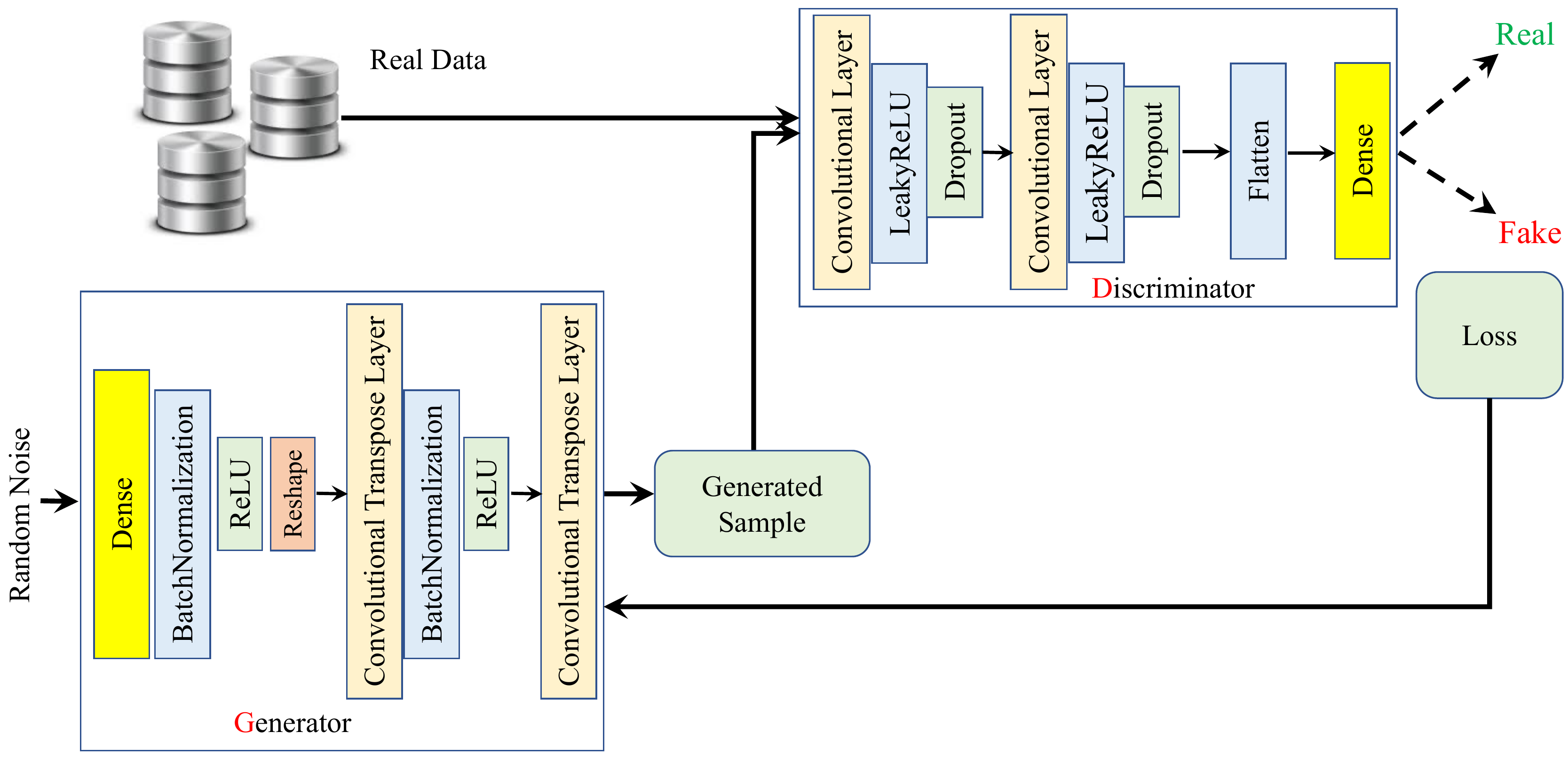}
\caption{\small GAN Architecture Used in GSelf-MDS }
\label{fig:GANarchitecture}
\end{figure}

Algorithm~\ref{alg:algorithm-GSelf-MDS} represents the GSelf-MDS pseudocode. In this algorithm, we use a \emph{while} to complete the data entry in the network, and it confirms the malware entry in real-time. The \emph{two} nested \emph{for} loops run the desired number of epochs for the \emph{generator} and \emph{discriminator} functions in GAN that generate adversarial samples. The for loop on lines ~\ref{lbl:for-G-start}-\ref{lbl:for-G-end} generates the fake samples $fk$ with the $\mathbb{G}$ function. After evaluation with the classification algorithm, we update the generator and discriminator functions. Focusing on lines~\ref{lbl:for-D-start}-\ref{lbl:for-D-end}, we consider the generated samples and the actual samples and use the $\mathbb{D}$ function for discrimination and consequently training $\mathbb{G}$ and $\mathbb{D}$. Fig.~\ref{fig:GANarchitecture} illustrates the structure of GAN used in this method. $par_{G}$, $par_{C}$ and $par_{dis}$ represent parameters of generator, discriminator and classifier ,respectively.

\begin{algorithm}[t]
	\footnotesize
\caption{\small GSelf-MDS Attack}
\label{alg:algorithm-GSelf-MDS}
\textbf{Input:}$X_{train}$,$par_{G}$, $par_{C}$, $par_{dis},G_{ep},D_{ep}$\\
\textbf{Output:} $X_{perturb}$
 
\begin{algorithmic}[1]
\color{black}
\footnotesize
\While{true}\label{lbl:while-do}
\If{$0 < length(G_{ep})$}
\For{each $i\in G_{ep}$}\label{lbl:for-G-start}
\State{$fk$ $\leftarrow$ Generate fake sample with $\mathbb{G}$}
\State{$C\{fk\} \leftarrow$ classify $fk$ by classifier $C$}
\State{Update $par_{{G}}$,$par_{{C}}$,$par_{{dis}}$ via $\mathbb{D}$ by gradient}
\EndFor \label{lbl:for-G-end}
\EndIf
\If{$0 < length(D_{ep})$}
\For{each $i\in D_{ep}$}\label{lbl:for-D-start}
\State{apply $\mathbb{G}$ to generate new $fk$}
\State{Pass $fk$ and real $rl$ samples to $C$} 
\State{Update $\mathbb{D}$ and $\mathbb{G}$} by  $C$ 
\EndFor \label{lbl:for-D-end}
\EndIf
\If{$length(G_{ep})==0$}
\State{\textbf{break}}
\EndIf
\EndWhile\label{lbl:while-end}
\State{$X_{perturb}= fk$}
\State{\textbf{return $X_{perturb}$}}
\end{algorithmic}
\end{algorithm}

\subsection{ASelf-MDS: Adversarial Self-Supervised Malware Detection system}\label{AdversarialSelf-Supervised} 
The third algorithm is termed as the Adversarial Self-Supervised Malware detection system or \emph{ASelf-MDS}. ASelf-MDS introduces adversarial strategies to preserve robustness in the data representations for downstream tasks with no need for their labels. ASelf-MDS enhances the applied fully supervised methods, and robustness in self-supervision against label corruption, adversarial examples, and universal input corruptions. Let $D_m$ and $T_m$ denote the unlabeled dataset and a malware detection task, respectively. In this context, the SSL method aims to produce $D_m$ and precisely classifies samples automatically. To this end, we form an optimization problem to minimise $L_p(c_p,c_{pf},D_m)$ as a pre-trained loss based on $c_p$ variants. Besides, $c_{pf}$ describes the extra parameters used for $T_m$. To generate a malware detection system and generate SSL training on it, we integrate \emph{three} malware perturbation attack mechanisms: Label Flipping Attack (LFA), Jacobian-based Saliency Map Attack (JSMA), and Fast Gradient Sign Method (FGSM) attack. Considering these attacks, our SSL model will be robust against the adversarial samples generated in the system. We explain the tuning and modifications of the applied attack mechanism on the SSL model as follows:

\begin{itemize}[leftmargin=*]
\item \textbf{LFA}\label{adversarialLFA}.LFA could alter the label of features and modify the interval of each sample in a cluster. LFA uses silhouette Clustering-based Label Flipping Attack (SCLFA) scheme to define Silhouette Value (SV) measurement. SV describes each cluster object's relation in comparison to other clusters and keeps it as normalized distribution range between [-1,1]. The value of SV is calculated based on the Euclidean distance method. When $SV=1$, the sample data is well-matched to its cluster. When $SV\leq 0$, the selected sample is a good candidate for label flipping~\cite{taheri2020defending}. The flipped label is defined as: 
\begin{equation}
\label{eq:eq-LFA}\small
    \mathcal{L}_k=
\begin{cases}
(x_k,y_k), & SV>0, \\
(x_k,|1-y_k|), & \text{otherwise}
\end{cases}
\end{equation}
where $\mathcal{L}_k$ be the label of the $k$-th sample.
\item \textbf{JSMA}\label{adversarialJSMA}. JSMA crafts adversarial examples adopting feed-forward neural network mechanism~\cite{papernot2016limitations}. We adopt the SoftMax function to normalize the output rate in our proposed malware detection system. Thus, the following equation holds:
\begin{equation}
\label{eq:eq-JSMA} 
\mathcal{G}_k(X)=\frac{e^{x_k}}{e^{x_0}+e^{x_1}},\ x_k=\sum_{l=1}^{f}w_{l,k}\cdot x_l+b_{l,k},
\end{equation}
where $\mathcal{G}(.)$ is a gradient function, $X$ is the input sample, and $f$ is the number of features. To build poison data to be injected to the datasets, as~\cite{papernot2016limitations}, we need to compute the Jacobian gradient ($J_F$) for each sample $X$ to predict the direction of perturbation, see Eq.~\eqref{eq:eq-step1}, and choose the maximum positive gradient in the perturbed models (see Eq.~\eqref{eq:eq-step2}). These activities continue until we reach the maximum number of changes or could successfully generate misclassification.

\begin{equation}
\label{eq:eq-step1} 
J_F\triangleq\frac{\partial F(X)}{\partial X}=\ \left[\frac{\partial F_k(X)}{\partial X_l}\right]_{k\in\{0,1\}, l\in[1,f]},
\end{equation}

\begin{equation}
\label{eq:eq-step2} 
k=\text{arg max}_{l\in[1,f], X_l=0}F_0(X_l).
\end{equation}

\item \textbf{FGSM}\label{adversarialFGSM}. FGSM~\cite{goodfellow2014explaining} could create untargeted, adversarial samples. Each FGSM sample is formed as:

\begin{equation}
 \label{eq:eqFGSM}\small
  \mathcal X_{adv} =X + \epsilon * \emph{sign( $\bigtriangledown_x \times J(\theta,x,y))$}, 
  \end{equation}
where $J(\theta,x,y))$ is the loss function and is the same as $L_p(c_p,c_{pf},D_m)$ in ASelf-MDS. To limit the level of perturbation ($X$ and $X_{adv}$) in the generated model, $\epsilon$ takes a binary value. We use the FGSM such that we first train a classification network and a separate auxiliary head. Then, we train this with the rest of the network to detect malware samples.

\end{itemize}

\begin{algorithm}[t]
	\footnotesize
\caption{\small ASelf-MDS Attack}
\label{alg:algorithm-ASelf-MDS}
\textbf{Input:} $X_{train}$,$step_{size}$, $T_{itr}$, $PerurbF(\cdot)$ \\
\textbf{Output:} $X_{perturb}$
\begin{algorithmic}[1]
\color{black}
\footnotesize
\State{$PerurbF(\cdot)\leftarrow LFA,JSMA,FGSM$}\label{lbl:ASelf-MDS-start}
\For{each $i$ in $T_{itr}$}\label{lbl:ASelf-MDS-for-start}
\State{$x_{i}$ $\leftarrow$ pass $x_{i}$ to $PerurbF(\cdot)$}
\State{Find gradients $p_t =\nabla_{x} \mathcal{L}(x_{i}, X_{train})$}
\State{compute perturbation using $x_{i+1} = x_i + \kappa \cdot sign (p_t)$}\label{lbl:ASelf-MDS-computeperturbation}
\State{Add perturbation to the $x_{i}$ using $X_{i+1} = clip(x_{i+1})$}\label{lbl:ASelf-MDS-addperturbation}
\State{$X_{perturb}=X_{perturb}+x_{i}$}
\EndFor\label{lbl:ASelf-MDS-for-end}
\State{\textbf{return $X_{perturb}$}} \label{lbl:ASelf-MDS-pass}
\end{algorithmic}
\end{algorithm}

Algorithm~\ref{alg:algorithm-ASelf-MDS} represents the ASelf-MDS pseudocode. In line~\ref{lbl:ASelf-MDS-start}, we adopt \emph{three} algorithms LFA, JSMA and FGSM to generate perturbed samples. Then, in a \emph{for} loop (lines~\ref{lbl:ASelf-MDS-for-start}-\ref{lbl:ASelf-MDS-for-end}), we repeat this process for a certain number of times. Each time we give the $x_i$ instance to the perturbation function $PerurbF(\cdot)$. We update the gradient $p_t$ based on the new instances for each time. Then, we generate the perturbation with the relation mentioned in line~\ref{lbl:ASelf-MDS-computeperturbation} and add it to the production dataset reported in line~\ref{lbl:ASelf-MDS-addperturbation}. Finally, we consider $x_{per}$ and pass it to the classification algorithm (line~\ref{lbl:ASelf-MDS-pass}).

\begin{algorithm}[t]
	\footnotesize
\caption{\small Self-Train Defence}
\label{alg:defalgorithm-Adv-self-Train}
\textbf{Input:} $X_{per}$, $Per_{B}$,$S_{size}$,$lr$,$itr$\\
\textbf{Output:} $X_{Corrected}$
 
\begin{algorithmic}[1]
\color{black}
\footnotesize
\State{$M \leftarrow$ self-supervised model}\label{lbl:def-m}
\State{Initialize $M$ parameters}\label{lbl:def-initial-m}
\For{each $ep$ in $itr$}\label{lbl:def-for1-start}
\For{each $Btch$ in $X_{per}$}\label{lbl:def-for2-start}
\State{$x_{adv}\leftarrow$ generate perturbation(seed=rand)}
\State{$z\leftarrow M(x_{adv})$}
\State{$y\leftarrow M(x_{per})$}
\State{recompute model parameters with gradient on $z$ and $y$}
\EndFor\label{lbl:def-for1-end}
\EndFor\label{lbl:def-for2-end}
\State{\textbf{$X_{Corrected} \leftarrow X_{per}+x_{adv}$}} \label{lbl:Adv-self-Train-pass}
\State{\textbf{return $X_{Corrected}$}} 
\end{algorithmic}
\end{algorithm}

\subsection{Defence: Adversarial Self
-Supervised Training}\label{AdversarialSelf-Defence}
In order to preserve the ML model against the aforementioned attack scenarios, we define adversarial self-supervised training method using the perturbed data as the \emph{self-train defence} method. Then, we add the newly generated recovered samples to the dataset to decrease the dataset's perturbation and help the ML return to the safe state. Algorithm~\ref{alg:defalgorithm-Adv-self-Train} presents the pseudocode of self-train defence method used against all three types of attacks. This defence method uses retraining based on perturbed data. In lines~\ref{lbl:def-m} and~\ref{lbl:def-initial-m} of this algorithm, we train the self-supervised model and initialize the steps. In lines~\ref{lbl:def-for1-start}-\ref{lbl:def-for2-end}, in \emph{two} nested \emph{for} loops, we generate new sample batches with gradient help and add them to the dataset. In this way, the model that is trained with this data will achieve higher accuracy.

\subsection{Time Complexity of Proposed algorithms}\label{TimeComplexity} 
The computational complexity of the attack and defence method is very crucial, especially for the real-time self-supervised training ML algorithm. Here, we analyze their computational complexities. Note that we dedicate an amount of the time to the training of discriminator and generator functions in all algorithms. Thus, the complexities of the proposed methods are listed below:

\begin{itemize}[leftmargin=*] 
\item \textbf{Self-MDS.} In this algorithm, we use a $while$ loop. Let $n$ be the number of samples; this loop's complexity is $\mathcal{O}(n)$. Another essential action that affects the complexity within this loop is a $for$ loop, which has complexity $\mathcal{O}(n)$.  Considering the computation time for Eq.~\eqref{eq:eq4} and the corresponding gradient, it is also in the order of $\mathcal{O}(n^2)$. To sum up, the algorithm's total complexity is in the order of $\mathcal{O}(n^2)$.

\item \textbf{GSelf-MDS Attack.} In this algorithm, we use a $while$ loop to generate adversarial samples. Let $n$ be the number of samples. We execute this loop as a percentage of the number of samples. Hence, this loop's maximum complexity is $\mathcal{O}(n)$. Inside this loop, we run two $for$ loops, which have a complexity of $\mathcal{O}(1)$, due to the fixed number of training generators and discriminator. In a nutshell, the complexity of the GSelf-MDS attack algorithm is in the order of $\mathcal{O}(n)$.

\item \textbf{ASelf-MDS Attack.} In this algorithm, the complexity mainly comes from running the $for$ loop. Let $n$ be the number of samples. This $for$ loop performs gradient operations for which the complexity is the order of $\mathcal{O}(n)$. Also, LFA, JSMA, and FGSM algorithms are used in this method, and $\mathcal{O}(max(LFA, JSMA, FGSM))$ should be added to the time complexity of this method. Therefore, the computational complexity of ASelf-MDS attack algorithm is in the order of $\mathcal{O}(n) + \mathcal{O}(max(LFA,JSMA, FGSM))$.

\item \textbf{Self-Train Defence} The complexity of the Self-Train algorithm is affected by \emph{two} nested $for$ loops. Let $n$ be the number of samples. In the worse case, both loops execute with $n$ samples. Hence, the self-train defence algorithm's computational complexity is in the order of $\mathcal{O}(n^2)$.
\end{itemize}

\section{Performance Evaluation}\label{resultAnalysis}
In this section, we evaluate the performance of our proposed attack and defence algorithms. We explain the simulation setup (\Cref{SimulationSetup}) and experimental results (\Cref{sec:results}) in the following subsections.

\subsection{Simulation Setup}\label{SimulationSetup}
Here, we explain the datasets (\Cref{sec:datasets}), simulation metrics (\Cref{sec:metrics}) and related settings (\Cref{sec:settings}).

\subsubsection{Datasets}\label{sec:datasets} 
To assess the proposed method on IoT data, we consider two IoT botnet datasets: IoT-23 dataset~\cite{parmisano2020labeled} and NBaIoT dataset~\cite{meidan2018n} which are explained in the sequel.
\begin{itemize}[leftmargin=*]
\item \emph{IoT-23 dataset~\cite{parmisano2020labeled}:}
This large dataset gathers the IoT's network traffic in the Stratosphere Laboratory, at the CTU University in the Czech Republic, funded by Avast Software in Prague, and published in January 2020. It includes twenty malware (infected IoT devices), and three benign (real/clean IoT devices) captures (pcap files) per device. To test the dataset, we implemented a specific malware sample on a Raspberry Pi and applied various actions. The traffic data is captured from three real/clean IoT devices: a Philips HUE smart LED lamp, a Somfy smart doorlock, and an Amazon Echo. 
The pcap files, include the benign and malware samples gathered from a consistent, controlled, network environment.

\item \emph{NBaIoT dataset~\cite{meidan2018n}:} This dataset mainly covers the real traffic samples gathered from nine IoT devices infected by Mirai and BASHLITE. With emphasis on the malicious data generation, they consider ten attacks and one benign sample category and form a multi-class classification strategy, which is infected through two botnets. The IoT devices trained with deep learning algorithm (i.e., autoencoder) over 66\% of the benign samples and form standard network traffic patterns. The remaining benign samples are merged with the malicious samples and implemented (deep) autoencoder as an anomaly detector in the testing phase.

\end{itemize}

\subsubsection{Metrics}\label{sec:metrics}
The proposed malware detection algorithms use the confusion matrix defined as follows:

\begin{itemize}[leftmargin=*]
\item \emph{True Positive ($\zeta$):} It indicates the number of correctly labelled samples that pertain to the class.
\item \emph{True Negative ($\eta$):} This explains the labelled samples, which are not on the class.
\item \emph{False Positive ($\iota$):} This describes the false classified samples referring to the class.
\item \emph{False Negative ($\omega$):} This explains the non-classified samples list.

\item \emph{$\Lambda$:} Accuracy explains as:
\begin{equation}
\label{eq:eq-accuracy}\small
\Lambda=\frac{\zeta+\eta}{\zeta+\eta+\iota+\omega}
\end{equation}

\item \emph{$\varphi$:} It presents the ratio of relevant samples to the retrieved samples:
\begin{equation}
\label{eq:eq-precision}\small
\varphi=\frac{\zeta}{\zeta+\iota}
\end{equation}

\item \emph{$\Re$:} It refers to recall and is computed as:
\begin{equation}
\label{eq:eq-recall}\small
\Re=\frac{\zeta}{\zeta+\omega}
\end{equation}
\item $\mathbb{F}_1$: It expresses the mean of precision and recall as:
\begin{equation}
\label{eq:eqf1}\small
\mathbb{F}_1=\frac{1}{\frac{1}{\Re}+\frac{1}{\varphi}}=2*\frac{\varphi*\Re}{\varphi+\Re}
\end{equation}

\item \emph{$\mathbb{FPR}$:} It is the ratio of negative samples wrongly classified as positive to the total number of actual negative samples. $\mathbb{FPR}$ is calculated as:
\begin{equation}
\label{eq:eq-fpr}\small
\mathbb{FPR}=\frac{\iota}{\iota+\eta}
\end{equation}
\item \emph{$\mathbb{AUC}$:} It describes the best model for predicting the class of events using all thresholds. It calculates the trade-off among misclassification percentage and $\mathbb{FPR}$ as:
 \begin{equation}
\label{eq:eq-auc}\small
\mathbb{AUC}=\frac{1}{2}\left(\frac{\zeta}{\zeta+\iota}+\frac{\eta}{\eta+\iota}\right)
\end{equation}
\end{itemize}

\subsubsection{System Setting} \label{sec:settings} 

The proposed self-supervised training attack and defence schemes perform three-phase of learning: training, validation and testings which consist of 60\%, 20\%, and 20\% of allocated samples for each dataset, simultaneously. We perform the experiments for IoT malware detection on an eight-core 4 GHz Intel Core i7 using 64-bit Windows 10 OS server, 16 GB of RAM, using Python 3.6.4. We adopt our method using Tensorflow (version 1.12.0) and Keras (version 2.2.4).

\begin{figure}[!htbp]
\centering 
\includegraphics[width=0.69\textwidth]{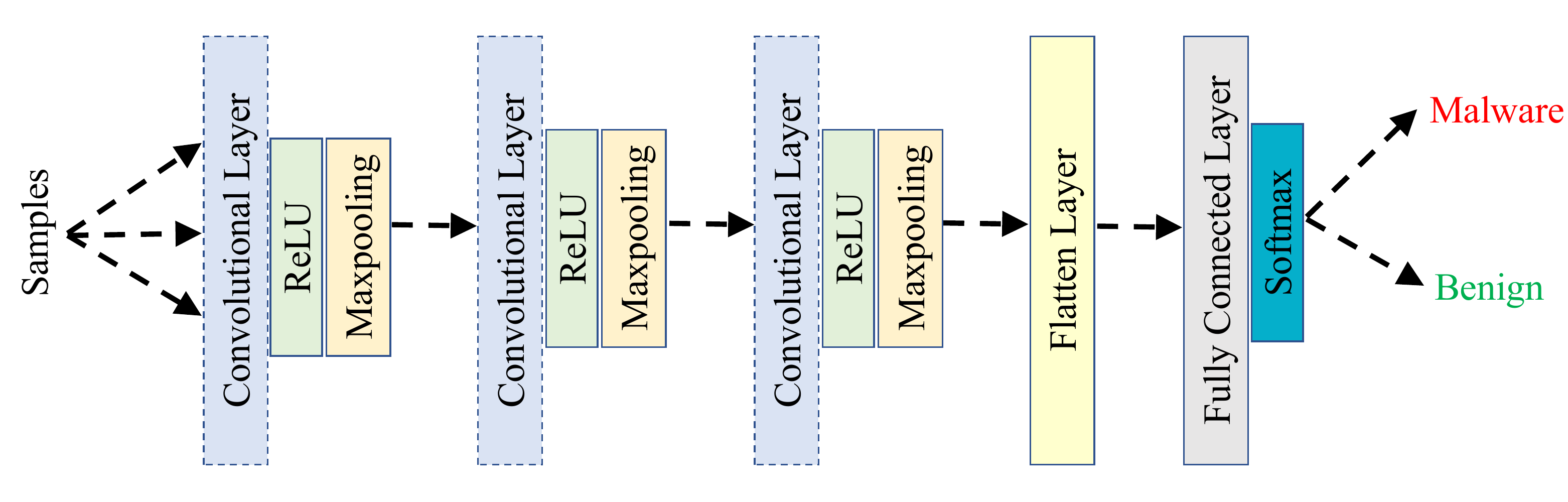}
\caption{\small The designed CNN classifier.}
\label{fig:cnn-classifier}
\end{figure}

Our convolutional neural network (CNN) architecture is depicted in Fig.~\ref{fig:cnn-classifier}. The designed \emph{CNN} is used as a classifier in all three proposed attacks. To design this \emph{CNN}, we apply sequential layers. We run three \emph{Convolutional} layers sequentially that use \emph{Maxpooling} and \emph{Relu} together. Flatten layer, fully connected layer and softmax function are the following parts of this CNN.

 \subsection{Results}\label{sec:results}
 
In this subsection, we assess the performance of the proposed attack and defence methods.

\begin{figure}[!htbp]
\begin{subfigure}[t]{0.45\textwidth}
	\centering
	\includegraphics[width=\textwidth]{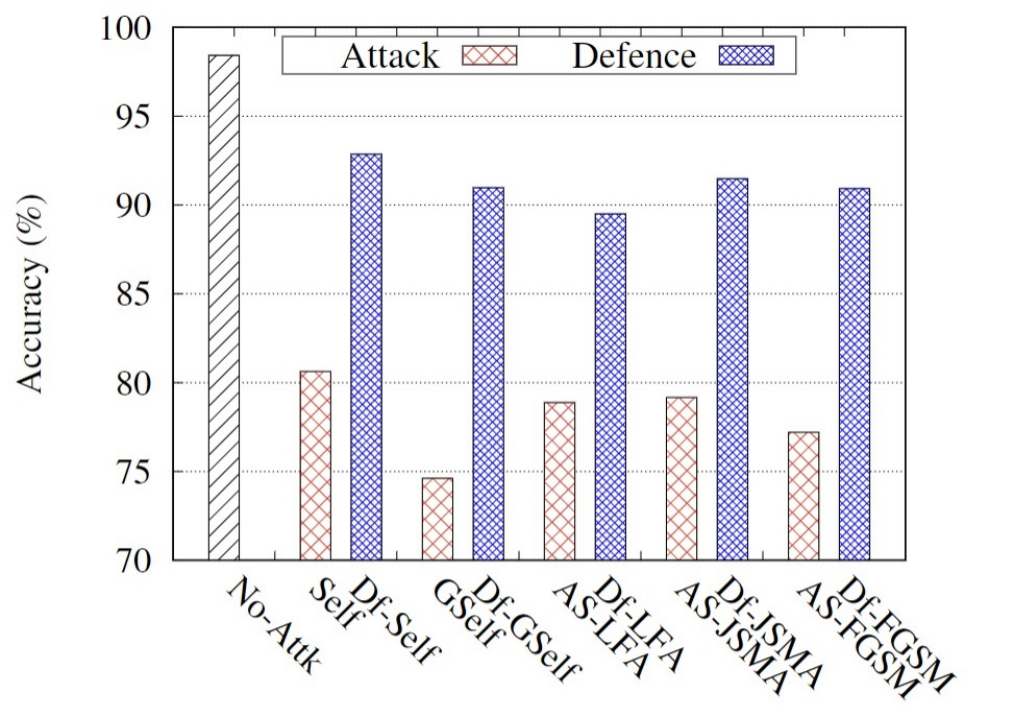}
	\caption{\small IoT23 dataset.}
	\label{fig:accuracy-IoT23}
\end{subfigure}
\begin{subfigure}[t]{0.45\textwidth}
	\centering
	\includegraphics[width=\textwidth]{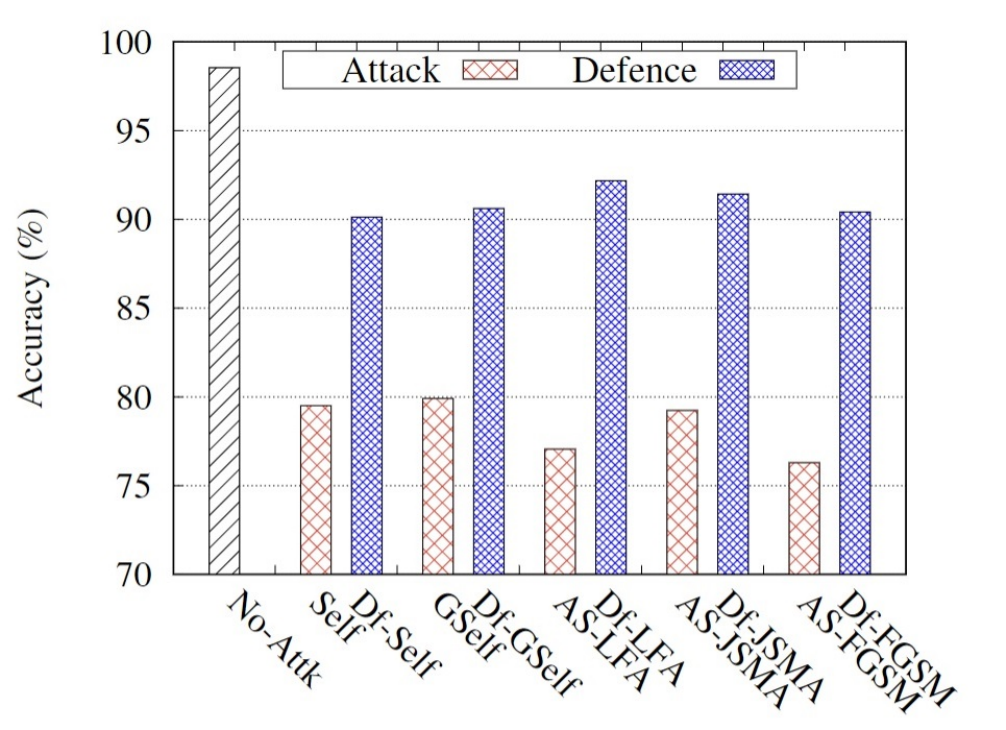}
	\caption{\small NBIoT dataset.}
	\label{fig:accuracy-NBIoT}
\end{subfigure}
\caption{\small The accuracy results of Self-MDS, GSelf-MDS, ASelf-MDS attack algorithms and the proposed self-train defence algorithms on different datasets. No-Attk:= without attack; DF:= defend; AS:= ASelf-MDS attack approach.}
	\label{fig:attack-comparison}
\end{figure}

\subsubsection{Comparing Methods based on Accuracy}\label{sec:accuracy-comparisons}
In Fig.~\ref{fig:accuracy-NBIoT}, we illustrate the accuracy results among the proposed attack and defence algorithms on the two datasets used in the paper. In Fig.~\ref{fig:accuracy-IoT23}, we show the result for the IoT23 dataset. We have the following findings:
(i) In the no-attack case, the learning model can classify the data with an accuracy of over 98\% (i.e., it is the same for both datasets). (ii) Right after applying the attack algorithms, we observe the trained model's accuracy decreases (see the red shaded bars in Fig.~\ref{fig:accuracy-NBIoT}). (iii) For the accuracy of the attack algorithm results, we realise that the most significant decrease in the accuracy results is related to applying the Gself-MDS attack, which is about 75\%. 

Interestingly, after running our proposed defence algorithm (i.e., adversarial self-supervised training algorithm) presented in Algorithm~\ref{alg:defalgorithm-Adv-self-Train}, we can increase the accuracy of the model created with the classification algorithm and this increment is more visible with the ASelf-MDS on FGSM perturbation methods. It confirms that our defence method could easily detect the untargeted adversarial samples added to the dataset and drop the newly added perturbed data from the affected dataset. 

Similarly, in Fig.~\ref{fig:accuracy-NBIoT}, we present the accuracy results for the NBIoT dataset. While applying the proposed attack algorithms, ASelf-MDS is the most effective attack algorithm with the average plunging of accuracy around 22\%. Also, Self-MDS is the second-best attack method with accuracy about 19.1\% and the least effective attack method in the list is GSelf-MDS with 18.7\%. On the other hand, by studying the defence solutions, the self-train defence algorithm running LFA technique can improve the accuracy around 16\% and is thus regarded as the best defence methods among the defence strategies.  

\begin{figure}[!htbp]
\begin{subfigure}[t]{0.45\textwidth}
	\centering
	\includegraphics[width=\textwidth]{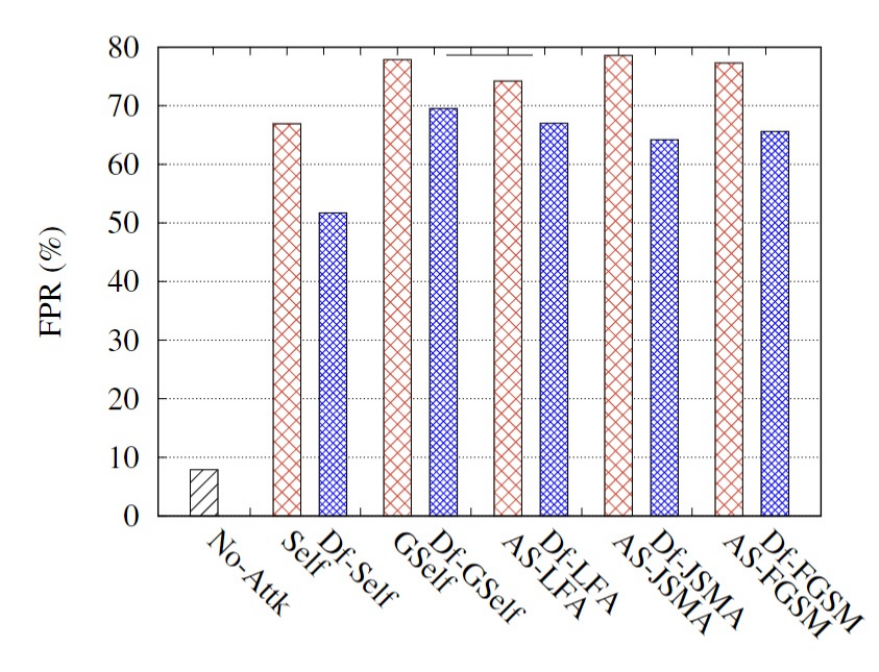}
	\caption{\small IoT23 dataset.}
	\label{fig:fpr-IoT23}
\end{subfigure}
\begin{subfigure}[t]{0.45\textwidth}
	\centering
	\includegraphics[width=\textwidth]{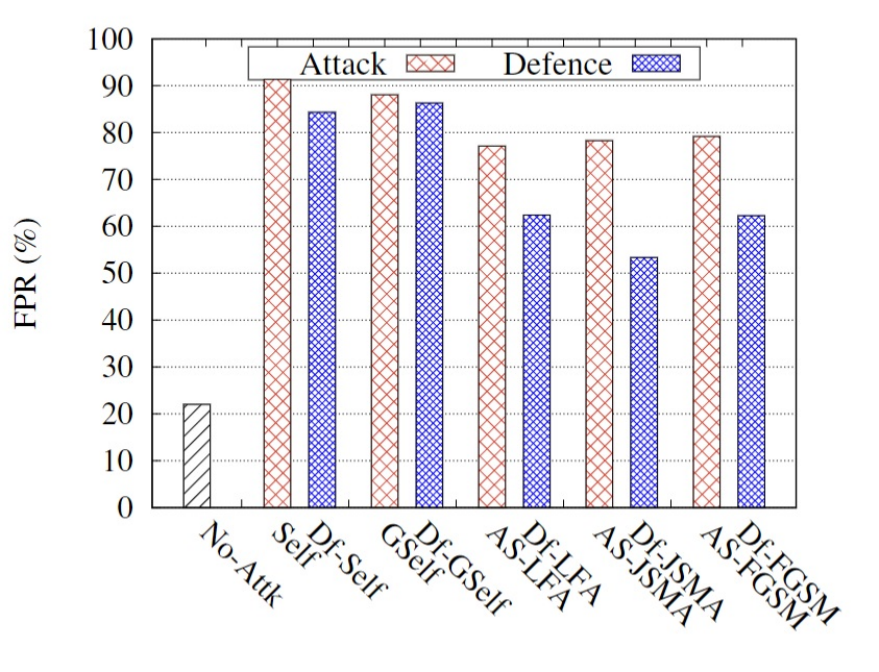}
	\caption{\small NBIoT dataset.}
	\label{fig:fpr-NBIoT}
\end{subfigure}
\caption{\small The FPR results of Self-MDS, GSelf-MDS, ASelf-MDS attack algorithms and the proposed self-train defence algorithms on different datasets. No-Attk:= without attack; DF:= defend; AS:= ASelf-MDS attack approach.}
	\label{fig:fpr-comparison}
\end{figure}
\begin{figure*}[!htbp]
\begin{subfigure}[t]{0.245\textwidth}
	\centering
	\includegraphics[width=\textwidth]{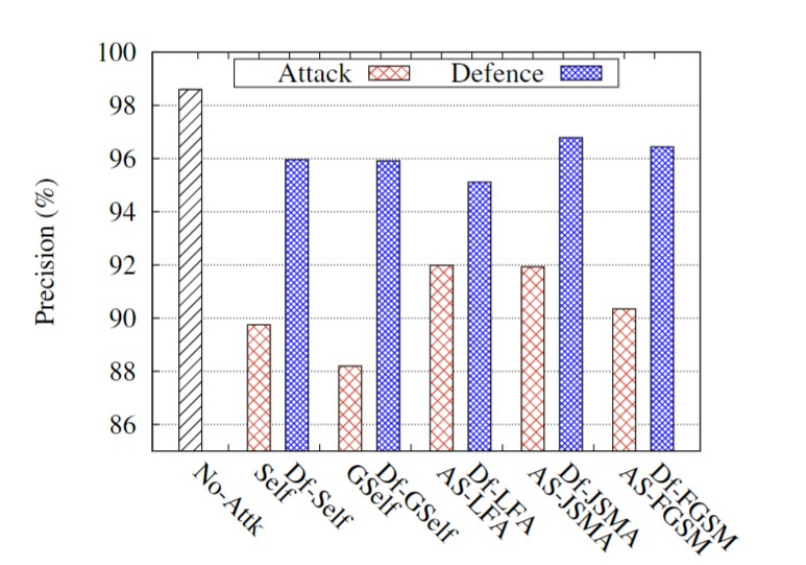}
	\caption{\small Precision-IoT23 dataset.}
	\label{fig:Precision-IoT23}
\end{subfigure}
\begin{subfigure}[t]{0.245\textwidth}
	\centering
	\includegraphics[width=\textwidth]{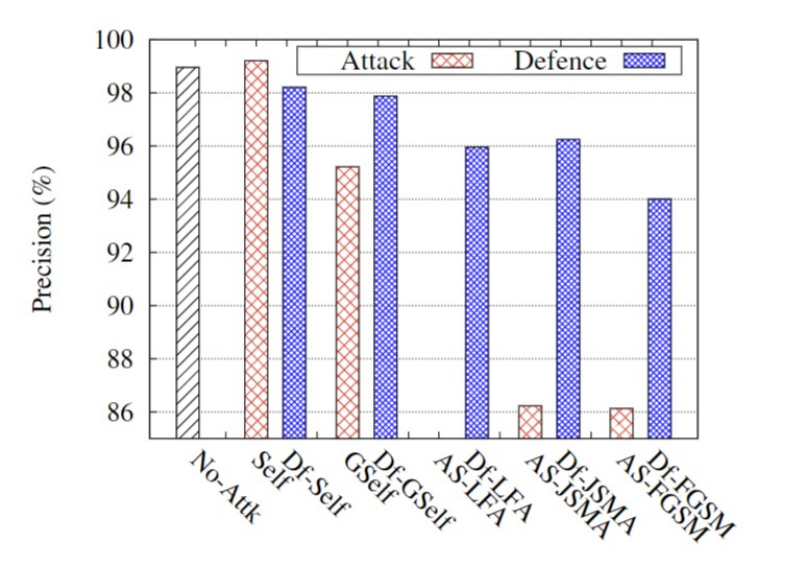}
	\caption{\small Precision-NBIoT dataset.}
	\label{fig:Precision-NBIoT}
\end{subfigure}
\begin{subfigure}[t]{0.245\textwidth}
	\centering
	\includegraphics[width=\textwidth]{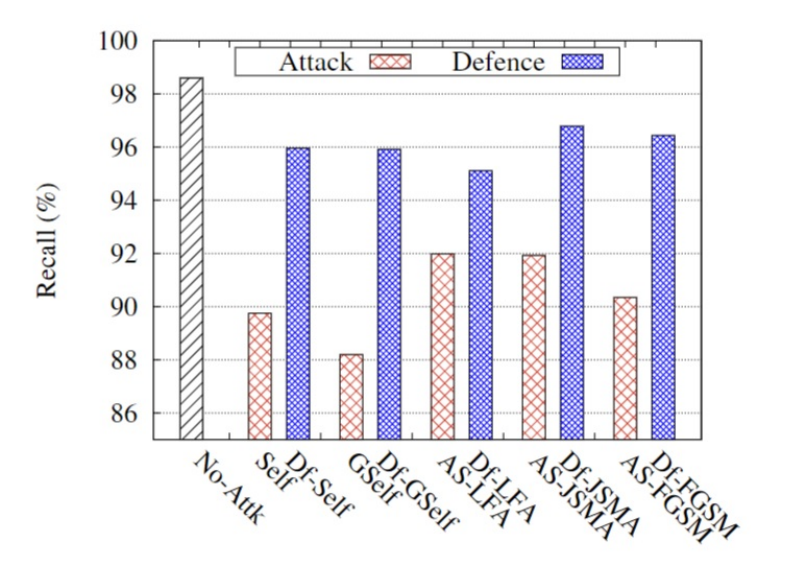}
	\caption{\small Recall-IoT23 dataset.}
	\label{fig:Recall-IoT23}
\end{subfigure}
\begin{subfigure}[t]{0.245\textwidth}
	\centering
	\includegraphics[width=\textwidth]{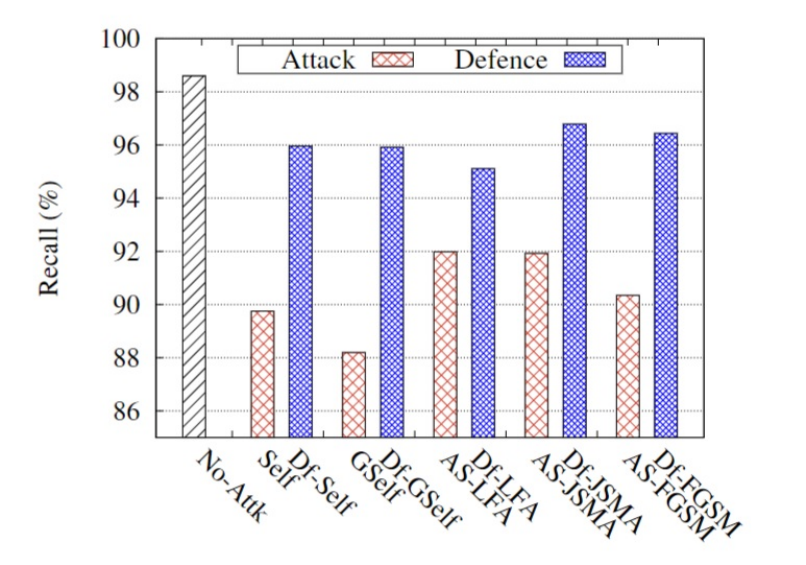}
	\caption{\small Recall-NBIoT dataset.}
	\label{fig:Recall-NBIoT}
\end{subfigure}
\caption{\small The precision ($\varphi$) and recall ($\Re$) results of Self-MDS, GSelf-MDS, ASelf-MDS attack algorithms and the proposed self-train defence algorithms on different datasets. No-Attk:= without attack; DF:= defend; AS:= ASelf-MDS attack approach.}
	\label{fig:precision-recall-comparison}
\end{figure*}
\subsubsection{Comparing Methods based on FPR}\label{sec:fpr-comparisons}
Fig.~\ref{fig:fpr-comparison} draws the FPR ratio among the attack and defence algorithms. In general, in adversarial ML, we must first check the FPR values and the corresponding accuracy values of each attack algorithm to analyse such an algorithm's destructiveness. In this way, we experience FPR values being low when an attack has not yet taken place. In the first step, a comparison of the two datasets shows that if the attack does not occur, the model obtains from the classification algorithm for the IoT23 dataset has an FPR value of less than 10\%. In contrast, in the NBIoT dataset, this value has reached over 20\%, which shows that despite the accuracy of 98\% (as demonstrated in Fig.~\ref{fig:attack-comparison}), the employed model is less successful. In the case of the IoT23 dataset, Fig.~\ref{fig:fpr-IoT23} shows that the FPR value significantly increases by the proposed attack methods. It means the model perturbed with these algorithms has not properly extracted benign samples from the whole sample set, leading to malware removal and causing a lot of False Positive $\iota$ values. Comparing the proposed attack algorithms, less success from the attacker's point of view is related to the self-supervised method, reversely, more success is given to the ASelf-supervised using the JSMA attack method. In this figure, applying the adversarial self-supervised defence method to the attacked data reduces the FPR values. Besides, using the defence method will not be able to produce a better model than when the attack does not take place. 

On the other hand, in the case of the NBIoT dataset, if we run the attack algorithm, it increases the FPR values on the ML model. In this dataset, the self-supervised and GSelf-supervised attacks are successful from the attacker's point of view. It has reached about 90\%, these differences in FPR values are probably due to the distribution of samples in the two datasets.

\subsubsection{Comparing Methods based on Precision/Recall}\label{sec:precision-recall-comparisons}

Fig.~\ref{fig:precision-recall-comparison} shows the two datasets' precision and recall values using the proposed attack algorithms and an adversarial self-supervised training defence algorithm. The concept of precision refers to the closeness of the measurements to each other. The data differ for different attack algorithms, confirming the results of the FPR values in  Fig.~\ref{fig:fpr-comparison}which shows that the amount of precision has increased after using the defence method.

From Figs.~\ref{fig:Precision-IoT23} to~\ref{fig:Recall-IoT23}, we reach \emph{three} main conclusions. First, after training and testing the final generated model for the attack and defence algorithm, we obtain similar values for the corresponding precision and recall for both datasets. It validates the preciseness of the generated ML model. Second, the GSelf-MDS attack adversely affects attack method and decreases the precision/recall by about 14\%. Finally, self-train defence using LFA perturbation is the most effective defence method that can influence precision/recall and raise this value to 96\%. The rationale is that when we increase the value of precision/recall, we boost the number of true positives $\iota$. However, it is always less than the case when no attack has ever taken place.

\subsubsection{Comparing Methods based on F1-score and AUC}\label{sec:f1-auc-comparisons}

Table~\ref{table:auc-f1} presents the $\mathbb{F}_1$ and $\mathbb{AUC}$ results of attack and defence algorithms for both datasets. From this table, it is observed that (i) when we run the attack algorithms the value of $\mathbb{F}_1$ and $\mathbb{AUC}$ reduces enormously and this rate affects on the model predication and increases the misclassification ratio and leads to having less $\mathbb{AUC}$. Focusing on IoT-23 dataset rows, GSelf-MDS attack algorithm is the most disruptive attack method that can diminish the $\mathbb{F}_1$ and $\mathbb{AUC}$ by around 15\% and 40\% compared with the case without attack state of the ML algorithm, respectively. Among the defence algorithms, self-train defence methods that applied on GSelf-MDS could recover the diminish rate most significantly comparing to the similar defence methods. Focusing on NAIoT dataset rows, Self-MDS's $\mathbb{AUC}$ rate decreases to half of its rate before applying an attack algorithm on the ML model, and this rate is the highest among the attack methods. Interestingly, the Self-Train defence algorithm that is designed to protect the ML model against ASelf-MDS is more robust than the other methods. Also, its $\mathbb{F}_1$ and the $\mathbb{AUC}$ recovering ratios are around 10\% and 15\%. 

\begin{table}[!htpb]
\caption{\small Comparison of $\mathbb{F}_1$ and $\mathbb{AUC}$ results between different solutions on two datasets. ST-Def:= Self-Train Defence.}
\label{table:auc-f1}
\begin{center}
\begin{adjustbox}{max width=\columnwidth}
\begin{tabular}{|l|l|c|c|}
\hline
\rowcolor[HTML]{EFEFEF} \textbf{Dataset}&\textbf{Algorithm}&\textbf{$\mathbb{F}_1$}&\textbf{$\mathbb{AUC}$}\\\hline
\multirow{11}{*}{\textbf{IoT-23}}&\textbf{Without Attack}&99.19&95.96\\\cline{2-4}\cline{2-4}
&\textbf{Self-MDS}&88.60&60.28\\\cline{2-4}
&\textbf{ST-Def (Self)}&96.17&72.34\\\cline{2-4}\cline{2-4}
&\textbf{GSelf-MDS}&84.67&51.78\\\cline{2-4}
&\textbf{ST-Def (GSelf)}&95.16&62.43\\\cline{2-4}\cline{2-4}
&\textbf{ASelf-MDS-LFA}&87.48&54.59\\\cline{2-4}
&\textbf{Self-Train Defence (LFA)}&94.28&63.24\\\cline{2-4}
&\textbf{ASelf-MDS-JSMA}&87.85&52.77\\\cline{2-4}
&\textbf{Self-Train Defence (JSMA)}&95.41&64.93\\\cline{2-4}
&\textbf{ASelf-MDS-FGSM}&86.46&52.80\\\cline{2-4}
&\textbf{Self-Train Defence (FGSM)}&95.09&64.09\\\hline\hline

\multirow{11}{*}{\textbf{NBIoT}}&\textbf{Without Attack}&99.26&88.76\\\cline{2-4}\cline{2-4}
&\textbf{Self-MDS}&88.34&44.14\\\cline{2-4}
&\textbf{Self-Train Defence (Self)}&94.72&53.56\\\cline{2-4}\cline{2-4}
&\textbf{GSelf-MDS}&88.55&47.34\\\cline{2-4}
&\textbf{Self-Train Defence (GSelf)}&95.02&53.00\\\cline{2-4}\cline{2-4}
&\textbf{ASelf-MDS-LFA}&86.64&56.00\\\cline{2-4}
&\textbf{Self-Train Defence (LFA)}&95.83&66.64\\\cline{2-4}
&\textbf{ASelf-MDS-JSMA}&88.11&55.88\\\cline{2-4}
&\textbf{Self-Train Defence (JSMA)}&95.30&70.50\\\cline{2-4}
&\textbf{ASelf-MDS-FGSM}&86.06&53.41\\\cline{2-4}
&\textbf{Self-Train Defence (FGSM)}&94.84&66.69\\\hline
\end{tabular}
\end{adjustbox}
\end{center}
\end{table}

\section{Conclusions and Future Directions}\label{conclusion}

Due to the lack of access to the IoT traffic sample labels in many cases, selecting self-supervised methods for detecting malware is needed. However, the presence of an adversary in the environment may attack models derived from such methods. In this paper, we presented self-supervised adversarial malware detection architecture called SETTI. In SETTI, we developed three types of adversarial attacks by self-supervised malware detection methods in an IoT environment and suggested a defence algorithm against them. We have examined the proposed methods on two benchmark datasets: IoT23 and NBIoT. We have observed that self-supervised methods are also vulnerable to adversarial attacks. Among the proposed methods, Self-MDS method is the most effective attack mechanism and it considers data entry and generation of adversarial samples in real-time. This work opens several avenues for future research. First, we aim to study entering on-line data and generation of on-line adversarial samples in self-supervised methods over the real-time traffic in the IoT network. Second, we plan to adopt cutting-edge self-supervised models for the discriminator and optimize them for data representations. Third, we can extend the proposed self-supervised GAN and enhance it to support fine-tune model representations. Finally, we can integrate our generated models with cutting-edge technologies such as self-attention, orthogonal normalization and regularization, to robustify performance in unconditional data traffic.

 
	\Urlmuskip=0mu plus 1mu\relax
    \bibliographystyle{ACM-Reference-Format}

    \bibliography{main.bib}
\end{document}